\documentclass[14pt]{extarticle}
\usepackage[utf8]{inputenc}
\usepackage{mathtools}
\usepackage{amsfonts}
\usepackage{amsmath}
\usepackage{amssymb}
\usepackage{fancyhdr}
\usepackage{hyperref}
\usepackage{braket}
\usepackage{bbm}
\usepackage{faktor}
\usepackage{tensor}
\usepackage{cleveref}
\usepackage{nicefrac}

\newcommand\be{\begin{equation}}
\newcommand\ba{\begin{eqnarray}}
\newcommand\ee{\end{equation}}
\newcommand\ea{\end{eqnarray}}
\numberwithin{equation}{section}

\usepackage[
   citestyle=numeric-comp,
   sorting=none,
   maxnames=10,
   giveninits=true,
  ]{biblatex}
 \renewbibmacro{in:}{}  %I am using this command to suppress the word 'In' which typically precedes a reference in biblatex%

\DeclareFieldFormat{postnote}{#1}
\DeclareFieldFormat{multipostnote}{#1} 
\DeclareFieldFormat{pages}{#1}
\title{Clifford group and stabilizer states from Chern-Simons theory}

\author{Howard J. Schnitzer\footnote{schnitzr@brandeis.edu} \\ Department of Physics \\ Brandeis University \\ Waltham, MA 02454}

\begin{document}
\flushbottom
\maketitle

\thispagestyle{fancy}
\renewcommand{\headrulewidth}{0pt}
%\rhead{BRX-TH 6635}

%  %
\begin{abstract}
The construction of generators of the Clifford group and of stabilizer states from Chern-Simons theory is presented for the Kac-Moody algebras $SU(2)_1$, $U(N)_{N,N(K+N)}$ with $N = 2$ and $K =1$, and $SU(N)_1$, extending results of Salton, et. al.
\end{abstract}

\clearpage

\section{Introduction}

We continue the study initiated by Salton, et. al. \cite{SaltonSwingleWalter2017} of entanglement from topology in Chern-Simons theory. This is a topological quantum field theory in an arena where the Euclidean path integral provides a map between geometry and states. For these three-dimensional quantum field theories there is a mapping $Z$, the functional integral, which relates the 3-dimensional manifold $\mathcal M$ to a probability amplitude $Z(\mathcal M )$. If $\mathcal M$ has a boundary, the boundary field configuration must be specified. That is, the path integral selects a state $\ket {\mathcal M}$ in a Hilbert space $\mathcal H_{\partial \mathcal M}$ associated to the boundary field configuration. If the boundary consists of several multiply connected components, $\mathcal H_{\partial \mathcal M} = \otimes^n_{i=1} \mathcal H_{\Sigma_i}$, where $\partial \mathcal M = \cup_{i=1}^n \Sigma_i$. This means that the different components $\Sigma_i$ are not coupled, so that the Euclidean path integral factorizes. 

In this paper we focus on the preparation of stabilizer states constructed from Chern-Simons theory defined on the n-torus Hilbert space.

In section 2 we discuss the Kac-Moody algebra $SU(2)_1$, while in section 3 it is shown that the unitary Kac-Moody algebra \cite{NaculichSchnitzer2007} $U(N)_{k,N(N+k)}$ for $N=2$, $k=1$ shares the conclusions of section 2. Therefore Chern-Simons theory for both $SU(2)_1$, and $U(2)_{1,6}$ allow one to prepare arbitrary stabilizer states in $\mathcal H_{T^2}^{\otimes_n}$. The Clifford group is then constructed from Clifford gates applied to $\ket{0}^{\otimes_n}$. The Choi-Jamiolkowski isomorphism then allows the preparation of an arbitrary element of the Clifford group \cite{Gottesman1998a,Gottesman1998b}.

In section 4 the analysis is extended to $SU(N)_1$, while in section 5 related issues are discussed.

\section{$SU(2)_1$} \label{sec:SU(2)_1}

The basis of the Kac-Moody algebra for $SU(2)_1$ is given in terms of Young tableau restricted to a single column, with basis $a=0,1$ where $a$ denotes the numbers of boxes of the tableau. We present the generators of the Clifford group for $SU(2)_1$ in terms of this basis, then the fusion matrix is
\begin{equation} \label{eq:2.1} N_{ab}^c \qquad \text{ with $a+b =c \mod 2$} \end{equation}
The modular transformation matrices are \cite{MlawerNaculichEtAl1991}, with standard normalization
\be \label{eq:2.2}
S_{ab} = \frac{1}{\sqrt 2} 
\begin{pmatrix}
1 & 1 \\ 1 & -1 
\end{pmatrix}, \quad a,b = 0,1 \ee
and \cite{NaculichRiggsSchnitzer1990}
\be
T_{ab} = \exp \left [2\pi i \left(h_a - \frac{c}{24}\right)\right]\delta_{ab}
\ee
where $c=1$ is the central charge for $SU(2)_1$ and the conformal dimension
\be h_a = \frac{C_2 (a)}{3} \ee
with $C_2 (a)$ the quadratic Casimir operator for the representation. In terms of angular momentum $j = a/2$, so that 
\be C_2 (a) = j(j+1) = \frac{a}{4}(a+2) \ee
thus
\be T_{ab} = \exp \left [2\pi i \left(\frac{a(a+2)}{12} - \frac{1}{24}\right)\right]\delta_{ab} \ee

The generators of the Clifford group are the Hadamard gate, the phase gate, and the controlled addition gate $c_{ADD}$, which satisfies \cite{Gottesman1998a,Gottesman1998b}
\be c_{ADD} \ket{a} \ket b = \ket a \ket{a + b \mod 2} \ee
Equation~\eqref{eq:2.2} shows that $S_{ab}$ is in fact the Hadamard gate: It is convenient to define $\omega = \exp i \pi$. Then in~\eqref{eq:2.2} 
\be \frac{1}{\sqrt 2} (\omega)_{ab} = \frac{1}{\sqrt 2} \begin{pmatrix} 1 & 1 \\ 1 & -1  \end{pmatrix} \ee
is the Hadamard gate. The phase gate $P_{ab}$ is
\be (\omega^{-1/12})\omega^{a(a+2)/6} T_{ab} = \begin{pmatrix} 1 & 0 \\ 0 & \omega \end{pmatrix} = P_{ab}. \ee
Given~\eqref{eq:2.1} and~\eqref{eq:2.2} we inherit the construction of Figure $3$ of Salton et al \cite{SaltonSwingleWalter2017} to construct the copy tensor, $c_{ADD}$, and a perfect tensor. Therefore for $SU(2)_1$ Chern-Simons theory one can prepare any stabilizer on the $n$-torus Hilbert space
\be \mathcal H_{T^2}^{\otimes_n} = ( \mathbbm{C}^2)^{\otimes_n} \ee

In section 5 we present this in a broader context.

\section{$U(N)_{K,N(K+N)}$ for $N=2, K=1$} \label{sec:U(N)_K_N(K+N)}

\subsection{}

In order to understand the special case $U(2)_{1,6}$, we first present the representations of the general case, which is described in detail
%of
in
section 2 of ref. \cite{NaculichSchnitzer2007}, and which is summarized here. The essential feature is that
\begin{equation}
    U(N)_{K,N(N+K)} = \faktor{\left[SU(N)_K \times U(1)_{N(K+N)} \right]}{\mathbb{Z}_N}
\end{equation}
which requires $K$ to be odd for consistency. Representations $(R,Q)$ of $SU(N)_K\times U(1)_{N(N+K)}$ must satisfy
\begin{equation} \label{eqn:Q=rModN}
    Q \equiv r \mod N
\end{equation}
where $r$ is the number of boxes of the Young tableau associated to $R$. There is an equivalence relation
\begin{equation}
    (R,Q) \simeq (\sigma(R), Q+N+K)
\end{equation}
where $\sigma$ is the simple current of $SU(N)_K$. Applying the simple current $N$ times, where $\sigma^N=1$, one obtains the equivalence
\begin{equation}
    (R,Q) \simeq (R, Q + N(N+K))
\end{equation}
so that $Q$ is restricted to the range
\begin{equation}
    0 \leq Q < N(N+K).
\end{equation}

The $U(N)$ representations $(R,Q)$ can be characterized by the extended Young tableau $\mathcal{R}$ with row lengths $\bar{l}_i \in \mathbb{Z}$, ($i=1$ to $N$). There is exactly one extended tableau $\mathcal{R}$ which satisfies
\begin{equation}
    0 \leq l_N \leq \ldots \leq l_1 \leq K.
\end{equation}
Hence the primary fields of $U(N)_{K,N(K+N)}$, where $K$ is odd, are in one to one correspondence with the Young tableaux $\mathcal{R}$ with no more than $N$ rows and $K$ columns. The number of such tableaux is $\binom{N+K}{N}$.

The modular transformation matrix for the $U(N)_{K,N(K+N)}$ character is \cite{NaculichSchnitzer2007}
\begin{equation} \label{eqn:S_AB}
    S_{AB} = \sqrt{\frac{N}{N+K}} S_{ab} \, e^{-\nicefrac{2\pi i Q_A Q_B}{N(N+K)}}
\end{equation}
where $S_{ab}$ is that of $SU(N)_K$. The subscripts $A$ or $a$ indicate whether one refers to $U(N)_{K,N(K+N)}$ or $SU(N)_K$. The modular transformation matrix for $\tau \to \tau + 1$ is \cite{NaculichSchnitzer2007}
\begin{equation}
    T_{AB} = \exp\left[ 2\pi i \left(h_A - \frac{c}{24}\right) \right] \delta_{AB}
\end{equation}
where the central charge
\begin{equation}
    c = \frac{N(NK+1)}{K+N}
\end{equation}
and
\begin{equation}
    h_A = \frac{\frac{1}{2}C_2(A)}{K+N}
\end{equation}
where $h_A = h(R,Q)$, and
\begin{subequations}
\begin{align}
    h(R,Q)  &= h_a + \frac{Q^2}{2N(N+K)} \\
            &= \frac{1}{2}\frac{C_2(R)}{(K+N)} + \frac{Q^2}{2N(N+K)} \\
            &= \frac{1}{2} \frac{C_2(R,Q)}{K+N}
\end{align}
\end{subequations}
and $Q=r \mod N$ from \eqref{eqn:Q=rModN}. Therefore
\begin{equation}
    T_{AB} = T_{ab} \; \exp\left[2\pi i \left( h_a + \frac{Q^2}{2N(N+K)} - \frac{c}{24} \right)\right] \delta_{AB}.
\end{equation}
The fusion matrix is
\begin{align}
    \tensor{N}{_{AB}^C} &= \sum_D \frac{S_{AD} S_{BD} (S_{CD})^{-1}}{S_{0D}} \\
    &= \tensor{N}{_{ab}^c} \sum_{Q_D} \exp \left[  -2\pi i \frac{\left(Q_A+Q_B-Q_C\right)Q_D}{N(N+K)} \right] \nonumber \\
    &= \tensor{N}{_{ab}^c} \delta_{(Q_A + Q_B - Q_C)} \nonumber
\end{align}
where
\begin{equation*}
    a+b = c \mod N,
\end{equation*}
together with \eqref{eqn:Q=rModN}, describes the fusion matrix $\tensor{N}{_{ab}^c}$, as well as $Q$ charge conservation by virtue of \eqref{eqn:Q=rModN}.

\subsection{$N=2, K=1$; $U(2)_{1,6}$}

Now specialize to $U(2)_{1,6}$, making use of the review of $U(N)_{K,N(K+N)}$ to discuss this case. The fusion matrix $\tensor{N}{_{ab}^c}$ satisfies $c = a+b \mod 2$, as does the fusion matrix of $SU(2)_1$, where $a,b,$ and $c$ are the number of boxes of a single column tableau. Then
\begin{equation}
    \tensor{N}{_A_B^C} = \tensor{N}{_a_b^c} \delta_{Q_A +Q_B -Q_C},
\end{equation}
where now $Q_A = a \mod 2$ from \eqref{eqn:Q=rModN}, so that charge conservation is automatically satisfied. Restricting \eqref{eqn:S_AB} to $N=2, K=1$, one can again inherit the construction of \textcite{SaltonSwingleWalter2017} to construct the $C_{ADD}$ gate. The Helmholtz gate and phase gate are essentially that of $SU(2)$, combined with $Q$ conservation.

\section{$SU(N)_1$} \label{sec:SU(N)_1}

Representations of $SU(N)_1$ are described by a single column tableau with $0 \leq N-1$ boxes. The fusion tensor is
\begin{equation}
    \tensor{N}{_a_b^c}, \quad \text{with} \quad a+ b = c \mod N.
\end{equation}
Therefore, this case closely parallels that of $SU(2)_1$ in \cref{sec:SU(2)_1}. The $C_{ADD}$ gate will satisfy
\begin{equation}
    C_{ADD} \ket{a}\ket{b} = \ket{a}\ket{a+b \mod N}
\end{equation}
so that with $S_{ab}$, $\tensor{N}{_a_b^c}$, and the phase gate, one constructs a basis for the Clifford group.

The modular transformation matrix normalized as in \cite{MlawerNaculichEtAl1991} is
\begin{equation}
    S_{ab} = \frac{(-i)^{N(N-1)/2}}{\sqrt{N}(N+1)^{(N-1)/2}} \det M(a,b)
\end{equation}
where $a,b = 0$ to $N-1$,
\begin{equation}
    M_{ij}(a,b) = \exp\left[ \frac{2\pi i \phi_i(a) \phi_j(b)}{N+1} \right],
\end{equation}
with $i=1$ to $N$ and
\begin{equation}
    \phi_i(a) = l_i(a) - i - \frac{r(a)}{N} + \frac{1}{2}(N+1)
\end{equation}
where
\begin{equation} \label{eqn:l_i}
    l_i =
    \begin{cases}
        \sum_{j=i}^{N-1} a_j & \text{for} \quad i = 1 \; \text{to} \; N-1 \\
        0 & \text{for} \quad i=N
    \end{cases}
\end{equation}
and $r(a) = \sum_{i=1}^{N-1}l_i(a)$ is the total number of boxes in the reduced Young tableau corresponding to the representation $a$.

The modular transformation matrix \cite{NaculichRiggsSchnitzer1990}
\begin{equation} \label{eqn:T_ab}
    T_{ab} = \exp\left\{ 2\pi i \left[ \frac{C_2(R)}{2(N+1)} - \frac{1}{24} \right] \right\} \delta_{ab}
\end{equation}
with the conformal dimension is
\begin{equation}
    h_a(R) = \frac{1}{2} \frac{C_2(R)}{N+1},
\end{equation}
with the quadratic Casimir operator
\begin{equation}
    C_2(R) = X + r(N+1) - \frac{r^2}{N}
\end{equation}
with $r(a)$ as above, and
\begin{equation}
    X = \sum_{i=1}^{N-1} l_i (l_i - 2i),
\end{equation}
where from \eqref{eqn:l_i}
\begin{equation} \label{eqn:l_i_order}
    l_1 \geq l_2 \geq \ldots \geq l_{N-1} \geq 0.
\end{equation}
Note $C_2(R)$ is quadric in $a$, vanishing when $a=0$.

Given $\tensor{N}{_a_b^c}$ with $c = a + b \mod N$, the detailed expression for $S_{ab}$ is not explicitly needed for the construction of Figure 3 of \textcite{SaltonSwingleWalter2017}. From \eqref{eqn:T_ab} to \eqref{eqn:l_i_order} one extracts an overall phase factor to obtain the phase gate
\begin{equation}
    P_{ab} = \exp (i \pi h_a (R)) \delta_{ab}.
\end{equation}
This, together with $C_{ADD}$ and $S_{ab}$, generates the Clifford group \cite{Gottesman1998a, Gottesman1998b}.

\section{Related issues}

In \cref{sec:SU(2)_1,sec:U(N)_K_N(K+N),sec:SU(N)_1} we generalized the results of Theorem 1 of \textcite{SaltonSwingleWalter2017} to $SU(2)_1$, $U(2)_{1,6}$ and $SU(N)_1$. The unifying feature which makes this possible is that the fusion tensors are all of the form
\begin{equation} \label{eqn:N_ab^c}
    \tensor{N}{_a_b^c} \; ; \quad a + b = c \mod N.
\end{equation}
This, together with the modular transformation matrices $T_{ab}$ and $S_{ab}$, allows one to construct the phase gate, and the $C_{ADD}$ gate, while $S_{ab}$ is (conjectured to be) an appropriate generalization of the Helmholtz gate. Thus, one can repeat the strategy of Figures 3 and 4, ff. of \textcite{SaltonSwingleWalter2017}. In particular, for $SU(N)_1$ the computation of Figure 4(a) gives $N^2$, that of Figure 4(b) yields $N^4$, etc., which means that the entanglement entropy of an arbitrary many torus system is
\begin{equation} \label{eq:52}
    S(A) = S(B) = S(C) = \log N.
\end{equation}
As a consequence the $SU(N)_1$ fusion tensor is equivalent to $g=1$ GHZ state, independent of $N$ for the states that can be distilled between $A$, $B$, and $C$ for $\partial M = A \cup B \cup C$ for an arbitrary tripartition of the boundary torii. Similarly, one should expect analogous results for $Sp(N)_1$ Chern-Simons theory since the fusion matrix satisfies \eqref{eqn:N_ab^c}.

It is known that a universal topological computer based on $SU(2)_K$ requires $K\geq 3$ \cite{Preskill}. This is exemplified by the work of \textcite{FreedmanLarsenWang2002a} which presents a detailed construction for $SU(2)_3$. Then level-rank duality shows that a universal topological quantum computer can be based on $SU(3)_2$ \cite{Schnitzer2018}. Level-rank duality then suggests that a universal topological quantum computer can be based on $SU(K)_2$, where $K \geq 3$.

Other applications of entanglement in Chern-Simons theory are discussed in refs. In particular, refs. \cite{DongFradkinEtAl2008,BalasubramanianFlissEtAl2017,BalasubramanianDeCrossEtAl2018,ChunBao2017,DwivediSinghEtAl2017} consider stabilizer states in $U(1)$ Chern-Simons theory.

In that context we follow \cite{BalasubramanianDeCrossEtAl2018}, where upper-bounds are derived for $SU(2)_K$, given an $n$-component link $\mathcal L^n \subset S^3$, and two sublinks $\mathcal L_A^m$ and $\mathcal L_{\bar A}^{n-m} \subset S^3$ such that a separating surface $\Sigma_{A | \bar A} \subset S^3$ is a connected, compact, oriented two-dimensional surface without boundary, where: $(1)$  $\mathcal L_A^m$ is contained in the handlebody inside $\Sigma_{A| \bar A}$, $(2)$ $\mathcal L_{\bar A}^{n-m}$ is contained in the handlebody outside $\Sigma_{A|\bar A}$, and $(3)$ $\Sigma_{A|\bar A}$ does not intersect any of the components of $\mathcal L^n$. Reference \cite{BalasubramanianDeCrossEtAl2018} presents a trivial upper-bound on the entanglement entropy for $SU(2)_K$, i.e.
\begin{equation} \label{eq:53}
S_{EE} (\mathcal L_A^m | \mathcal L_{\bar A}^{n-m}) \leq \ln (K+1) \min (m, n-m) 
\end{equation}
and a tighter upper bound
\begin{equation} \label{eq:54}
S_{EE} (\mathcal L_A^m | \mathcal L_{\bar A}^{n-m}) \leq \ln \left [ \sum_{u=0}^K \frac{1}{S_{0u}^{2\min (g_\Sigma) -2}} \right ] 
\end{equation}
Specialize to $SU(2)_1$, where $S_{00} = 1/ \sqrt{2}$ and $S_{01} = -1/\sqrt{2}$, so that \eqref{eq:53} becomes
\begin{equation}
S_{EE} (\mathcal L_A^m | \mathcal L_{\bar A}^{n-m}) \leq (\ln 2) \min (m, n-m),
\end{equation}
while \eqref{eq:54} for $\min (g_\Sigma) \geq 1$ becomes
\begin{equation}
S_{EE} (\mathcal L_A^m | \mathcal L_{\bar A}^{n-m}) \leq (\ln 2) \min (g_\Sigma)
\end{equation}
Further for $SU(N)_1$, recalling \eqref{eq:52} we expect
\begin{equation}
S_{EE} (\mathcal L_A^m | \mathcal L_{\bar A}^{n-m}) \leq (\ln N ) \min (g_\Sigma). 
\end{equation}
\section{Acknowledgements}

We thank Isaac Cohen, Alastair Grant-Stuart, and Andrew Rolph for assistance in the preparation of the paper.

\printbibliography

%\bibliographystyle{JHEP}
%\bibliography{references.bib}

\end{document}